\documentclass [pre, twocolumn, showpacs] {revtex4}

\begin{document}

\title{The nature of most probable paths at finite temperatures}

\author{Pratip Bhattacharyya}
 \email{pratip@cmp.saha.ernet.in}

\affiliation {Theoretical Condensed Matter Physics Division,
  Saha Institute of Nuclear Physics,
  Sector - 1, Block - AF, Bidhannagar, Kolkata 700 064, India}

\date{March 28, 2003}

\begin{abstract}

We determine the most probable length of paths at finite temperatures,
with a preassigned end-to-end distance and a unit of energy assigned to
every step on a $D$-dimensional hypercubic lattice.
The asymptotic form of the most probable path-length shows a
transition from the directed walk nature at low temperatures to the
random walk nature as the temperature is raised to a critical value
$T_c$. We find $T_c = 1/(\ln 2 + \ln D)$. Below $T_c$ the most probable
path-length shows a crossover from the random walk nature for small
end-to-end distance to the directed walk nature for large end-to-end
distance; the crossover length diverges as the temperature approaches 
$T_c$. For every temperature above $T_c$ we find that there is a
maximum end-to-end distance beyond which a most probable path-length
does not exist.

\end{abstract}


\maketitle

\indent Among the different paths that can be drawn on a lattice, two
of the most common are generated by random walks and directed walks.
The length $n$ of a path is measured as the total number of steps from
one end to the other, where each step covers a unit lattice bond.
The paths are classified according to the relation between their
lengths $n$ and the distance $r$ between their endpoints.
Paths generated by a random walk are characterised by $n \propto r^2$
\cite{Chandrasekhar1943, Reif1965, Haus1987} whereas paths generated
by a directed walk have $n \propto r$~\cite{Halpin-Healy1995}.
The problem of determining the relation between
$n$ and $r$ under a given condition (such as the minimum
energy condition for the path) can be defined in two ways~: either
to determine the end-to-end distance $r$ for paths of a preassigned
length $n$, or, to determine the length $n$ of paths with a preassigned
end-to-end distance $r$. The problem of the first type is realised
in linear polymers seeking a minimum energy configuration on a
substrate on in a solvent~\cite{Gennes1979, Barat1995, Witten1998},
whereas the problem of the second type is only of geometric interest.

\indent In this paper we address the problem of the second type.
Consider a $D$-dimensional hypercubic lattice. Let us assume that
every step on the lattice (i.e., to travel along one lattice bond)
requires a unit of energy. Now consider two points on the lattice
separated by a distance $r$~:

\begin{equation}
r^2 = \sum_{i = 1}^D r_i^2,
\label{eq:end-to-end}
\end{equation}

\noindent where $r_i$ is the component of $\bf r$ in the $i$-th
dimension. The length $n$ of any path connecting these two points may
be written as~:

\begin{equation}
n = \sum_{i = 1}^D n_i,
\label{eq:path-length}
\end{equation}

\noindent where $n_i$ is the number of lattice bonds along the path
in the $i$-th dimension. The paths of minimum energy connecting the
two points are the paths of shortest length between the two points~:

\begin{equation}
n_{\rm min} = \sum_{i = 1}^D \left \vert r_i \right \vert.
\label{eq:path-length-min}
\end{equation}

\noindent This may be considered as the most probable path-length
at zero temperature. Now our aim is to determine the most probable
path-length $n_{\rm mp} (r, T)$ at finite temperatures between the
two given points separated by a distance $r$. Temperature represents
noise. At $T = 0$ there is no noise; therefore the only posible
path-length between the two given points is that of the shortest
course from one point to the other. At finite temperatures noise
raises the value the most probable path-length above the minimum.
The most probable paths at finite temperatures are not the minimum
energy paths, but are paths of minimum free energy.

\indent We first calculate the number $N(n, r)$ of distinct paths of
length $n$ on the lattice, connecting two points separated by a
distance $r$. The number of distinct ways of assigning $n_i$ steps
to the $i$-th dimension when $n_1$, $\ldots$ , $n_{i-1}$ steps
have been already assigned to $i-1$ other dimensions is given
by the binomial coefficient

\begin{equation}
\left ( \begin{array}{c}
                 n + n_i - \sum_{k=1}^i n_k \\
                 n_i
        \end{array} \right ) .
\label{eq:factor1'}
\end{equation}

\noindent Thus the number of distinct ways of assigning
$n_1$, $\ldots$ , $n_D$ steps to dimensions $1$, $\ldots$ , $D$
respectively is given by the polynomial coefficient~:

\begin{equation}
\prod_{i=1}^D \left ( \begin{array}{c}
                               n + n_i - \sum_{k=1}^i n_k \\
                               n_i
                      \end{array} \right )
 = { n! \over \prod_{i=1}^D n_i !} ,
\label{eq:factor1}
\end{equation}

\noindent and the number of distinct ways in which $n_i$ steps
can be distributed among the positive and the negative directions
in the $i$-th dimension such that the difference between their
numbers is $r_i$ is given by the binomial coefficient

\begin{equation}
\left ( \begin{array}{c}
                 n_i \\
                 {n_i - r_i \over 2}
        \end{array} \right ) .
\label{eq:factor2}
\end{equation}

\noindent Therefore

\begin{eqnarray}
N(n, r) & = & \sum_{\{n_i\}} \: { n! \over \prod_{i=1}^D n_i !}
              \prod_{i = 1}^D \left ( \begin{array}{c}
                                               n_i\\
                                               {n_i - r_i \over 2}
                                      \end{array} \right ) \nonumber \\
 & = & \sum_{\{n_i\}}
       {n! \over \prod_{i = 1}^D 
                  \left ( {n_i - r_i \over 2} \right ) !
                  \left ( {n_i + r_i \over 2} \right ) ! }.
\label{eq:path-number}
\end{eqnarray}

\noindent The final expression in the above equation is obtained by
expanding the binomial coefficients. The summation indices $\{ n_i \}$
are subject to the conditions of Eq.~(\ref{eq:path-length}) and
Eq.~(\ref{eq:path-length-min}).

\indent At a finite temperature $T$ the statistical weight of a
step along any path is $\exp(- 1 / T)$, since a step on the lattice
requires a unit of energy. Therefore the partition function for
paths of length $n$ and end-to-end distance $r$ is given by~:

\begin{equation}
G_T(n, r) = N(n, r) \: \exp(-n / T),
\label{eq:partition-function-n}
\end{equation}

\noindent and the partition function for all paths with a
preassinged end-to-end distance $r$ at the temperature $T$
is given by~:

\begin{equation}
G_T(r) = \sum_{n = n_{\rm min}}^\infty G_T(n, r).
\label{eq:partition-function}
\end{equation}

\noindent For large path-lengths $n$ and $n_i \gg r_i$, we can use
Stirling's approximation~\cite{Graham1994} for the factorials in
Eq.~(\ref{eq:path-number})~:

\begin{eqnarray}
n! & \sim & \sqrt{2 \pi n} \: n^n \: \exp(-n), \\
\left ( {n_i \pm r_i \over 2} \right ) ! & \sim &
                     \sqrt{2 \pi \left ( {n_i \pm r_i \over 2} \right ) }
                     \: \left ( {n_i \pm r_i \over 2} \right )
                        ^{\left ( {n_i \pm r_i \over 2} \right )}
                          \nonumber \\
             & & \times \exp \left ( - \: {n_i \pm r_i \over 2} \right ) .
\label{eq:stirling}
\end{eqnarray}

\noindent Consequently Eq.~(\ref{eq:partition-function-n}) may be
written as~:

\begin{eqnarray}
G_T(n, r) & \sim & {1 \over (2 \pi)^{D - 1/2}} \sum_{\{ n_i \}}
 \exp \left [ n \ln n - \sum_{i = 1}^D n_i \ln n_i \right . \nonumber \\
    & & \hspace{1.0cm} + \frac{1}{2} \ln n - \sum_{i = 1}^D \ln n_i
        + (n + D) \ln 2 \nonumber \\
    & & \hspace{2.0cm} \left . - \sum_{i = 1}^D {r_i^2 \over 2 n_i}
        - {n \over T} \right ].
\label{eq:partition-function-n-approx1}
\end{eqnarray}

\noindent For large path-lengths $n$, the most probable value of
$n_i$, as well as the mean of $n_i$, is equal to $n/D$ for all
dimensions $i$. Hence we assume~:

\begin{equation}
n_i \approx \langle n_i \rangle = {n \over D}.
\label{eq:mp-mean-n_i}
\end{equation}

\noindent The partition function in
Eq.~(\ref{eq:partition-function-n-approx1}) now reduces to the form~:

\begin{eqnarray}
G_T(n, r) & \sim & {1 \over (2 \pi)^{D - 1/2}} \: \exp \left [
        - \left ( D - \frac{1}{2} \right ) \ln n \right . \nonumber \\
   & &  \hspace{2.0cm} + (n + D) (\ln 2 + \ln D) \nonumber \\
   & &  \hspace{2.0cm} - \left . {D \over 2 n} r^2 - {n \over T}
        \right ].
\label{eq:partition-function-n-approx2}
\end{eqnarray}

\noindent For the most probable value $n_{\rm mp}$ of the path-length
at temperature $T$, the free energy $- T \ln G_T(n, r)$ is minimum~:

\begin{equation}
\left [ - T {{\rm d} \over {\rm d} n}
              \ln G_T(n, r) \right ]_{n = n_{\rm mp}} = 0,
\label{eq:partition-f-at-n_mp}
\end{equation}

\noindent which leads to the following condition~:

\begin{equation}
2 \left [ \left ( \ln 2 + \ln D \right ) - {1 \over T} \right ] n_{\rm mp}^2
 - (2 D - 1) n_{\rm mp} + D r^2 = 0.
\label{eq:n_mp-condition}
\end{equation}

\noindent Solving the above equation we obtain the most probable
path-length as a function of the end-to-end distance and the
temperature~:

\begin{equation}
n_{\rm mp} (r, T) = {1 - \sqrt{1 - {8D \over (2D - 1)^2}
                        \left ( {1 \over T_c} - {1 \over T} \right ) r^2}
              \over
              {4 \over 2D - 1} \left ( {1 \over T_c} - {1 \over T} \right )}
\label{eq:n_mp}
\end{equation}

\noindent where
\begin{equation}
{1 \over T_c} = \ln 2 + \ln D .
\label{eq:T_c}
\end{equation}

\noindent In Eq.~(\ref{eq:n_mp}) only the negative sign of the
square-root is admissible, in order to ensure that $n_{\rm mp}$ is
always positive. In the simplest case of a one-dimensional lattice it
was found in an earlier study that $T_c = 1/ \ln 2$~\cite{Bhattacharyya2001}.
The result for $D = 1$ can now be obtained directly from the
general expression of $T_c$ given in the equation above.

\indent At low temperatures, i.e. $T < T_c$, the expression for the
most probable path-length in Eq.~(\ref{eq:n_mp}) may be written as~:

\begin{equation}
n_{\rm mp} (r, T) = {2D \: r_\times^2 \over 2D - 1}
 \left [ \sqrt{1 + \left ( {r \over r_\times} \right ) ^2} - 1 \right ]
\label{eq:n_mp-at-T<T_c}             
\end{equation}

\noindent where
\begin{equation}
r_\times (T) = {2D - 1 \over 2 \sqrt{2D}}
 \left ( {1 \over T} - {1 \over T_c} \right ) ^{-1/2}
\label{eq:end-to-end-crossover}
\end{equation}

\noindent is a characteristic end-to-end distance depending only
on the temperature and the dimension of the lattice. The significance
of this distance will be evident from the following analysis.
For end-to-end distances which are much smaller than $r_\times$
the most probable path-length given in Eq.~(\ref{eq:n_mp-at-T<T_c})
reduces to the form~:

\begin{equation}
n_{\rm mp} \approx {D \over 2D - 1} \: r^2, \hspace{1.0cm} r \ll r_\times
\label{eq:n_mp-r<<rx}
\end{equation}

\noindent which has the feature of a random walk. For end-to-end
distances which are much larger than $r_\times$,
Eq.~(\ref{eq:n_mp-at-T<T_c}) reduces to the form~:

\begin{equation}
n_{\rm mp} \approx \sqrt{D \over 2} \left ( {1 \over T} - {1 \over T_c}
 \right ) ^{-1/2} r, \hspace{1.0cm} r \gg r_\times
\label{eq:n_mp-r>>rx}
\end{equation}

\noindent which has the feature of a directed walk.
Thus $r_\times (T)$ is a crossover distance which marks the crossover
in the nature of the most probable path-length from the random 
walk class to the directed walk class at any temperature $T < T_c$.
Eq.~(\ref{eq:end-to-end-crossover}) shows that the crossover
distance diverges as the temperature is raised to $T_c$ which may be
formally identified as a critical point. The asymptotic form (i.e.,
for large $r$) of the most probable path-length undergoes a transition
from the directed walk nature at $T < T_c$ to the random walk nature
at $T_c$. The limiting form of the most probable path-length as
the temperature approaches $T_c$ may be obtained directly from
Eq.~(\ref{eq:n_mp})~:

\begin{equation}
n_{\rm mp} = {D \over 2D - 1} \: r^2, \hspace{1.0cm} T \to T_c.
\label{eq:n_mp-at-T_c}
\end{equation}

\noindent This result for $T \to T_c$ is true for all values of $r$.

\indent For high temperatures, i.e. $T > T_c$, the most probable
path-length in Eq.~(\ref{eq:n_mp}) may be written as~:

\begin{equation}
n_{\rm mp} (r, T) = {2D \: r_{\rm max}^2 \over 2D - 1}
 \left [ 1 - \sqrt{1 - \left ( {r \over r_{\rm max}} \right ) ^2} \right ] .
\label{eq:n_mp-at-T>T_c}             
\end{equation}

\noindent where
\begin{equation}
r_{\rm max} (T) = {2D - 1 \over 2 \sqrt{2D}}
 \left ( {1 \over T_c} - {1 \over T} \right ) ^{-1/2}
\label{eq:end-to-end-max}
\end{equation}

\noindent is another characteristic end-to-end distance depending only
on the temperature and the dimension of the lattice. At any temperature
$T > T_c$, Eq.~(\ref{eq:n_mp-at-T>T_c}) shows that for end-to-end
distances $r > r_{\rm max}(T)$ a real and finite value of the most probable
path-length does not exist. For end-to-end distances much smaller than
$r_{\rm max}(T)$ Eq.~(\ref{eq:n_mp-at-T>T_c}) reduces to the form~:

\begin{equation}
n_{\rm mp} \approx {D \over 2D - 1} \: r^2, \hspace{1.0cm} r \ll r_{\rm max}
\label{eq:n_mp-r<<r_max}
\end{equation}

\noindent which has the same random walk feature found in
Eq.~(\ref{eq:n_mp-r<<rx}) and Eq.~(\ref{eq:n_mp-at-T_c}).
\noindent Eq.~(\ref{eq:end-to-end-crossover}) and
Eq.~(\ref{eq:end-to-end-max}) show that the two characteristic
end-to-end distances below and above $T_c$, i.e. $r_\times$ and
$r_{\rm max}$, diverge in the same way as $T \to T_c$.

\indent In this paper we have studied a temperature driven
transition in the geometry of most probable paths on a
$D$-dimensional hypercubic lattice. In the low temperature phase
$(T < T_c)$ the most probable paths show the appearence of a
random walk for small end-to-end distances but eventually cross
over to the form of a directed walk for large end-to-end distances.
Close to the critical temperature $(T \to T_c)$ the most probable
paths are like random walks for all end-to-end distances. In the
high temperature phase $(T > T_c)$ the very existence of most
probable paths is restricted by a maximum end-to-end distance for
each temperature. The critical temperature is found to be
$T_c = 1 / ( \ln 2 + \ln D )$ which gets lower with increasing
dimension of the lattice because of a simple reason~: the larger
the dimension, the greater the entropy of the paths, and therefore,
the lower is the noise (given as the temperature) required to bring
the random walk nature.

\begin{acknowledgments}

I am grateful to Bikas K. Chakrabarti and D. Stauffer for their comments
on the manuscript.

\end{acknowledgments}

\end{document}